\documentclass[a4paper,11pt]{article}
\pdfoutput=1 

\usepackage{jheppub2} 
\usepackage{bbm,bm,graphicx,mathtools,color,hyperref,slashed}
\usepackage{amssymb,amsmath}
\usepackage{extarrows}
\usepackage{comment}
\usepackage[all]{xy}
\usepackage[dvipsnames]{xcolor}

\graphicspath{{./Figures/}}

\def\Z{{\mathbb Z}}
\def\R{{\mathbb R}}

\preprint{.}

\title{Towards Stabilization on Noncommutative Torus }

\author[1]{Canberk G\"uvendik,}

\affiliation[1]{Department of Physics, North Carolina State University, Raleigh, NC 27607, USA}

\emailAdd{canberkgvndk06@gmail.com}

\abstract{Recent introduction of center vortices with 't Hooft flux on two torus compactification leads to a new semiclassical regime where confinement is analytically calculable \cite{Tanizaki:2022ngt}. In this work, we investigate the stability of the classical minima for gauge fields under quantum corrections.  Although the classical  $\Z_N \times \Z_N$ symmetric minima is stable at small-$N$, because of the nature of the quantum corrections, it can be destabilized at sufficiently large-$N$.   Using Morita equivalence, we switch to field theory on noncommutative torus instead of working with theory on torus with 't Hooft flux in a certain limit. Noncommutative Yang-Mills theory compactified on two torus leads to a tachyonic instability and it leads to the spontaneous breaking of translation symmetry.  We discuss that spontaneous breaking of translation symmetry is identical to the $\Z_N \times \Z_N$ center symmetry breaking similar to an older version of  
large N  twisted Eguichi-Kawai model. 
We compute the photon polarization diagram for noncommutative  U(1) theory in the presence of adjoint fermions and show that they stabilize the tachyonic instability on the noncommutative torus and restore the broken symmetry.    }

\begin{document}
\maketitle

\section{Introduction}\label{sec:introduction}
There are two weakly coupled regimes of Yang-Mills theory and QCD adiabatically connected to the strongly coupled regime, and furthermore,   using semiclassical methods, confinement, chiral symmetry breaking, and multi-branch structure of the vacua can be calculated analytically. One of them is the theory formulated on $\R^3\times S^1$ with suitable deformations where confinement is due to magnetic bions \cite{Unsal:2007jx,Poppitz:2021cxe,Unsal:2008ch} and the other one is theory formulated on $\R^2 \times T^2$ with 't Hooft flux (twisted boundary conditions) where confinement is achieved with center vortices \cite{Tanizaki:2022ngt}. In this work, we focus on the latter and investigate a subtle issue that appears in large-$N$ limit. 
When Yang-Mills theory formulated on $\R^2 \times T^2$ with twisted boundary conditions and torus compactified on $x^1-x^2$ directions, Polyakov loops on $T^2$ do not commute with each other and they satisfy \cite{Tanizaki:2022ngt}
\begin{eqnarray}
P_1 P_2=P_2 P_1 e^{(2\pi i n_{12}/N)}
\label{Polyakov}
\end{eqnarray}
where $n_{12}\in \Z_N$ and is known as 't Hooft flux. This algebra can be satisfied and the classical vacuum can be given by clock and shift matrices (see eq.\eqref{clock}) such that \newline $P_1=S$,$P_2=C$. Since $tr(P_1^{n_1}P_2^{n_2})=0$ for any $(n_1,n_2)\neq (0,0)$ mod N, the $\Z_N ^{[0]}\times \Z_N ^{[0]}$ symmetry is unbroken classically. For fixed-N, 
these vacua are stable at sufficiently weak coupling. 
However, it is not certain whether quantum fluctuations destabilize this classical vacua or not at a fixed-size torus as $N$ is dialed. In fact, there may be non-commutativity of limits between small $L$ and large-$N$.  The problem is related to the realization of symmetry breaking and volume independence in the twisted Eguichi Kawai model \cite{GonzalezArroyo:1982hz,GonzalezArroyo:2010ss,Teper:2006sp}. On the other hand, there exists Morita equivalence on $T^2$ such that a  U(N) with 't Hooft flux is dual to a noncommutative U(1) gauge theory \cite{Guralnik2,Szabo}. Intriguingly, even a U(1) noncommutative theory is asymptotically free \cite{Barbon}, and the same beta function as non-abelian theory appears. Noncommutative theories also show interesting properties such as UV/IR mixing \cite{Minwalla}. Then, one can ponder about the equivalence of $\Z_N \times \Z_N$ symmetry breaking on the noncommutative theory. It turns out that there exists a tachyonic instability \cite{Bietenholz,ArmoniLopez} arising at the one-loop perturbation theory and it leads to spontaneous breaking of translation invariance on the noncommutative side. It is interesting to see the instability arising from quantum corrections that can dominate the tree-level term. The instability on two torus can be written in terms of fractional momentum modes $\kappa$ and the corresponding electric flux $e_i=p\epsilon_{ij}\kappa _j$ where $p$ depends on noncommutativity parameter through the relation $\theta=\frac{p}{N}$ such that \cite{Guralnik}
\begin{eqnarray}
E^2=\frac{1}{R_{nc}^2}(\vec{\kappa}^2-g^2f(\frac{\vec{e}}{N}))
\label{instabilitygeneral}
\end{eqnarray}
where $R_{nc}$ is the radius of noncommutative torus. The instability would occur when $E^2<0$. For small values of $\vec{e}$,  $f(\frac{\vec{e}}{N})) \approx \frac{N^2}{\vec{e}^2}$, as we will see directly from \eqref{tachyonic} and \eqref{instability}.  Therefore, using 't Hooft coupling $\lambda =g^2 N$, the dispersion relation can be written as $\frac{1}{R_{nc}^2}(\vec{\kappa}^2-\frac{\lambda N}{\vec{e}^2})$. Indeed, at fixed N, there is no instability at small $\lambda$(small $L$ by asymptotic freedom). But at fixed small $\lambda$(small $L$), an instability would especially emerge for large N.  
\par
To our knowledge, there are two ways to overcome this instability: a special choice of $n_{\mu \nu}$, the ' t Hooft twist which depends on N \cite{Perez,2+1,Fibonacci,LargeNTEK,Perez2} or including matter fields in the adjoint representation \cite{Bietenholz}. Especially, it is known that the inclusion of adjoint matter fields stabilizes the center symmetry breaking and restores the large-N volume independence in $\R^3\times S^1$ theories \cite{Kovtun-Unsal,Poppitz:2021cxe}. Therefore, it is natural to think that the addition of adjoint fermions would stabilize the symmetry breaking on $T^2 \times \R^2$ as well. In this work, we focus on the latter solution and show that the inclusion of adjoint fermions indeed stabilizes the tachyonic instability in $T^2 \times \R^2$.

\section{Conventions and Morita Duality}\label{sec:Morita}
\label{sec:ConventionsandMoritaDuality}
We will follow the conventions given in \cite{Guralnik}. Noncommutativity is only present on the two torus for the compactification $T^2 \times \R^d$. D is taken as total dimensions whereas d is taken as the number of noncompact and commuting directions. We will have the torus on 1-2 plane and $[x^i,x^j]=2\pi i R^2 \theta \epsilon ^{ij}$ is taken where $\theta$ is dimensionless noncommutativity parameter. 
\par

Instead of working on the noncommutative torus, we will work on the commutative torus with deformed algebra. The way to achieve this is through the Moyal star product. We will replace all multiplication symbols with star products. We define the Moyal star product of two functions \cite{Guralnik,Szabo,Barbon}
\begin{eqnarray}
f(x)*g(x)=e^{\pi i R^2 \theta \epsilon _{ij} \partial _{x}^i \partial _{y}^j}f(x)g(y)|_{y\rightarrow x}
\label{Moyal}
\end{eqnarray}

On the other hand, Morita duality is an SL(2,Z) duality \cite{Guralnik3,Alvarez}. If an element of SL(2,Z) is $\begin{pmatrix}
p & q \\
r & s 
\end{pmatrix}$, then ps-qr=1 should be satisfied. Moreover, one can choose generators of the SL(2,Z) group as $S=\begin{pmatrix}
0 & -1 \\
1 & 0 
\end{pmatrix}$  ,$T=\begin{pmatrix}
1 & -1 \\
0 & 1 
\end{pmatrix}$\cite{GapedPhases}. 
According to duality, the noncommutative U(1) theory is dual to the commutative U(N) gauge theory. For rational $\theta=\frac{p}{N}$, the conversion table for the variables is as follows \cite{Guralnik2}:
\begin{table}
\begin{center}
\begin{tabular}{ |c|c| } 
 \hline
 U(1) NC & U(N) C\\
 \hline
 $\theta=-\frac{p}{N}$ & $\theta=0$\\
 \hline
$R^2$ & $\frac{R^2}{N^2}$  \\ 
 \hline
 $\phi =0$ & $\phi=-\frac{m}{N}$  \\ 
 \hline
\end{tabular}
\end{center}
\caption{Conversion table for Morita dual theories}
\end{table}
where $\phi$ is denoted as the 't Hooft flux, background field for the $Z_N$ center symmetry \cite{Unsal:2020yeh}. \par 
On noncommutative torus for $U=e^{i\frac{x_1}{R}}$,$V=e^{i\frac{x_2}{R}}$ one can show that \cite{Guralnik} 
\begin{eqnarray}
    U \ast V= V \ast U e^{-2\pi i \theta }
    \label{holonomync}
\end{eqnarray}. In general, mode expansions on noncommutative torus can be done using the \par $U^{k_1}\ast V^{k_2}e^{-\pi i \theta k_1 k_2}=e^{i\frac{\vec{x}}{R}.\vec{k}}$ for instance, for a scalar field $\phi (\vec{x})=\sum _{\vec{k}\epsilon Z^2} \phi _{\vec{k}} e^{i\frac{\vec{x}}{R}.\vec{k}}$. \par 
Instead of working with scalar fields on the noncommutative torus, one can work with matrix-valued fields satisfying relation (2.2) on the usual torus. These matrices are NxN clock and shift matrices given as follows: \cite{Guralnik}
\begin{eqnarray}
  C =
  \begin{bmatrix}
    1 & & \\
     & e^{2\pi i  \frac{p}{N}} & \\
    & \ddots & \\
    & & e^{2\pi i p \frac{(N-1)}{N}}
  \end{bmatrix}
 S =
  \begin{bmatrix}
    0 & 1 & \\
     & 0 & 1 &\\
    & &\ddots & \ddots \\
    1& & 
  \end{bmatrix}
\label{clock}
\end{eqnarray}
and $CS=SCw$ with $w=e^{2\pi i  \frac{p}{N}}$.

\section{Rederivation of the instability}
\label{sec:Rederivationofinstability} 

We will rederive the instability obtained from the gluon polarization diagrams mentioned in \cite{Guralnik}. One can reach Feynman rules for noncommutative U(N) gauge fields from ordinary SU(N) gauge theory by replacing \cite{Barbon,Szabo,Alvarez}:
\begin{equation}
f^{abc} \rightarrow f^{abc} cos(\frac{p_a \times p_b}{2})+d^{abc} sin(\frac{p_a \times p_b}{2})
\label{fromSU(N)}
\end{equation}
where we defined  $p_a \times p_b $ =$p_a \theta^{ab} p_b$. For U(1) gauge group and $d^{000}=2$ normalization, we will substitute $f^{abc} \rightarrow 2 sin(\frac{p_a \times p_b}{2})$.  We have the diagrams given in Figure 1 to compute.\par
\begin{figure}[h]
\begin{center}
\includegraphics[width=0.9\textwidth]{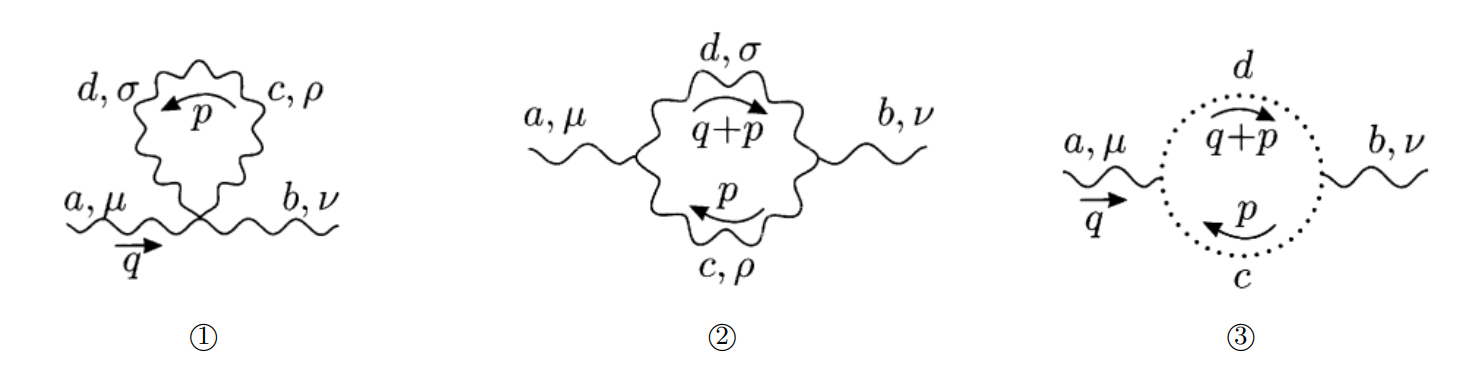} 
\end{center}
\caption{Gluon polarization diagrams from \cite{PeskinSchroeder199508}. Dots represent the ghost lines.}
\label{fig:stage}
\end{figure}
Let us write the self-energy obtained from the diagrams first in the notation of Peskin-Schroeder then we will convert it to Guralnik's notation \cite{Guralnik}. 

\begin{equation}
\nonumber
\textcircled{1}=-\frac{g^2}{2}\int \frac{d^4 p}{(2\pi)^4} \frac{g_{\rho \sigma} \delta ^{cd}}{p^2} [2f^{ace}f^{bde}(g^{\mu \nu} g^{\rho \sigma}-g^{\mu \sigma}g^{\nu \rho})] 
\end{equation}
We need to rewrite $\textcircled{1}$ in the following form
\begin{equation}
\nonumber
\textcircled{1}=-\frac{g^2}{2}\int \frac{d^4 p}{(2\pi)^4} \frac{g_{\rho \sigma} \delta ^{cd}(p+q)^2}{p^2 (p+q)^2} [2f^{ace}f^{bde}(g^{\mu \nu} g^{\rho \sigma}-g^{\mu \sigma}g^{\nu \rho})] 
\end{equation}

\begin{eqnarray}
\begin{gathered}
\nonumber
\textcircled{2}=-\frac{g^2}{2}\int \frac{d^4 p}{(2\pi)^4} \frac{f^{acd}f^{bcd}}{p^2 (p+q)^2} [N^{\mu \nu}]  \\
N^{\mu \nu}=[g^{\mu \rho}(q-p)^\sigma +g^{\rho \sigma}(2p+q)^\mu +g^{\sigma \mu}(-p-2q)^\rho]\otimes[\delta ^\nu _\rho (p-q)_\sigma +g_{\rho \sigma}(-2p-q)^\nu +\delta ^v _\sigma (p+2q)_\rho]
\end{gathered}
\end{eqnarray}

\begin{equation}
\nonumber
\textcircled{3}=-\frac{g^2}{2}\int \frac{d^4 p}{(2\pi)^4} \frac{f^{cbd}f^{dac}}{p^2 (p+q)^2} (-2(p+q)^\mu p^\nu)
\end{equation}

Using D as the total dimension and taking $D=d+2$ with Euclidean metric we will gather terms inside the integrals while arranging structure constants in the correct order
\begin{eqnarray}
\begin{gathered}
\textcircled{1}_{in}= -2(p+q)^2 g^{\mu \nu} (d+1) \\
\textcircled{2}_{in}= g^{\mu \nu}[2p^2+5q^2+2pq]+(d+2)\left(q^\mu q^\nu  +4 p^\mu p^\nu +2(p^\mu q^\nu+p^\nu q^\mu)\right)\\-\left(6q^\mu q^\nu +6 p^\mu p^\nu +3(p^\mu q^\nu +p ^\nu q^\mu)\right) \\
\textcircled{3}_{in}=-2(p+q)^\mu p^\nu
\end{gathered}
\label{Insideofintegrals}
\end{eqnarray}
where previously in $\textcircled{3}$ indices of the structure constants were ordered differently compared to $\textcircled{1}$ and $\textcircled{2}$ which resulted in an extra minus sign. \par
To convert these to Guralnik's notation, replace the loop momenta as $p_\mu = K_\mu$ whereas inflowing momenta is denoted as $q_\mu=-P_\mu$. For torus, we take $K_\mu =(k_\alpha , \frac{l_i}{R})$ and $P_\mu =(p_\alpha , \frac{n_i}{R})$ where $\alpha$ index runs from 1 to d, number of non-compact dimensions, i index runs from 1,2 for compact directions. Rewriting \eqref{Insideofintegrals} in this notation
\begin{eqnarray}
\begin{gathered}
\textcircled{2}_{in}= g^{\mu \nu}[2K^2+5P^2-2KP]+d\left(P^\mu P^\nu  +4 K^\mu K^\nu -2(K^\mu P^\nu+K^\nu P^\mu)\right)\\-4 P^\mu P^\nu +2 K^\mu K^\nu -(K^\mu P^\nu +K ^\nu P^\mu) \\
\textcircled{1}_{in}= -2 g^{\mu \nu} (d+1)(K^2+P^2-2KP) \\
\textcircled{3}_{in}=-2K^\mu K^\nu +2K^\nu P^\mu 
\end{gathered}
\label{InsideofintegralsGuralnik}
\end{eqnarray}
Adding the terms in \eqref{InsideofintegralsGuralnik} and using the fact that the final expression is symmetric under the exchange of $\mu \leftrightarrow \nu$ we obtain the same expression they use as $f_{\mu \nu}$ in the text.
\begin{equation}
f_{\mu \nu}=-g_{\mu \nu}[2d(K^2+P^2-2KP)-3P^2-2KP]+(d-4)P_\mu P_\nu +4d K_\mu K_\nu -2d(K_\mu P_\nu +K_\nu P_\mu)
\end{equation}
Finally using the phase factors coming from compact directions after replacing the structure constants we have $\sin{(\frac{l_i \times  n_j}{2 R^2})}^2$. Converting this to $\frac{1}{2}-\frac{1}{2}\cos{(\frac{l_i \times  n_j}{ R^2})}$. We will be interested in the nonplanar part from the cosine part and $\frac{1}{2}$ part is the planar part. The extra minus sign coming from in front of the cosine part will be incorporated in the quantum inverse propagator (see \eqref{quantuminversepropagator}). \par 
First, recall that while replacing the structure constant, we got a factor of 4 which is canceled by $\frac{1}{2}$ in the denominator of $\frac{g^2}{2}$ and $\frac{1}{2}$ coming from $\cos$ part of the nonplanarity. Secondly, recall that $l_i \times  n_j=2\pi R^2 \theta l_i n_j \epsilon ^{ij}$. After all these considerations, we can finally write the polarization tensor for the nonplanar part on the noncommutative torus.
\begin{equation}
\Pi _{\mu \nu}=\frac{g^2}{(2\pi R)^2}\sum _{\vec{l}\epsilon Z^2} \int \frac{d^d k}{(2\pi)^d} \frac{\cos{(2\pi \theta l_i n_j \epsilon _{ij})}}{K^2 (K-P)^2} f_{\mu \nu}
\end{equation}

From now on we will take $R=1$ to compute easier. At the end of the calculation, we will reintroduce the R dependence. At this point, we will start the main calculation. We start with introducing the Feynman parameters:
\begin{equation}
\frac{1}{K^2 (K-P)^2}=\int _0 ^1 \frac{dx}{[(K-P)^2x+K^2(1-x)]^2}=\int _0 ^1 \frac{dx}{[K^2-2KPx+P^2x]^2}
\label{Feynmanparameters}
\end{equation}
In this derivation, as the first important step, we will separate the loop momenta into noncompact and compact directions. We will achieve this by writing $K_\mu \rightarrow K_\mu +L_\mu $ where this time $K_\mu =(k_\alpha,0)$ and $L_\mu =(0,l_i)$, our new $K_\mu$ will be different than the original one. The denominator in \eqref{Feynmanparameters} becomes in the new notation
\begin{equation}
\int _0 ^1 \frac{dx}{[K^2-2KPx+P^2x]^2}=\int _0 ^1 \frac{dx}{[K^2+L^2-2KP_1x-2LP_2x+P^2x]^2}
\label{Feynmandenominator}
\end{equation}
where we also separated the inflowing momenta in noncompact and compact directions. We introduced $P_1=(p_\alpha ,0)$,$P_2=(0,n_i)$. Notice that in the new definition, K and L are orthogonal. \par 
Now we will introduce the Schwinger parameters for the equation \eqref{Feynmandenominator}
\begin{equation}
\int _0 ^1 \frac{dx}{[K^2+L^2-2KP_1x-2LP_2x+P^2x]^2} =\int _0 ^1 dx \int _0 ^\infty d\alpha \; \alpha e^{-\alpha ([K^2+L^2-2KP_1x-2LP_2x+P^2x])}
\end{equation}
For this expression we would like to integrate out the loop momenta in noncompact directions after all, we do not expect the final result to depend on them since we are interested in EFT on torus. To make the noncompact loop momenta part Gaussian, we will shift the momenta there such that $S=K-P_1 x$ and integrand $dk=ds$. 
\begin{equation}
    \int _0 ^1 dx \int _0 ^\infty d\alpha \; \alpha e^{-\alpha ([S^2+L^2-2LP_2x+P_1^2 x(1-x)+P_2^2x])}
\end{equation}
\par Now, we need to decompose the $f_{\mu \nu}$ in the same way. We will separate the momenta again. First, separate $K_\mu$
\begin{equation}
\begin{split}
\begin{gathered}
f_{\mu \nu}=4dK_\mu K_\nu +4d(K_\mu L_\nu +K_\nu L_\mu)+4dL_\mu L_\nu -2d(P_\mu K_\nu +P_\nu K_\mu)\\ -2d(P_\mu L_\nu +P_\nu L_\mu)+(d-4)P_\mu P_\nu\\ -g_{\mu \nu}(2dK^2+2dL^2-(4d+2)KP-(4d+2)LP+(2d-3)P^2)
\end{gathered}
\end{split}
\end{equation}
Now, separate the inflowing ones ($P_1$ and $P_2$).
\begin{equation}
\begin{split}
\begin{gathered}
f_{\mu \nu}=4dK_\mu K_\nu+4d(K_\mu L_\nu+K_\nu L_\mu)-2d(P_{1\mu}K_\nu+P_{1\nu}K_\mu)-2d(P_{2\mu}K_\nu+P_{2\nu}K_\mu)\\-g_{\mu \nu}(2d K^2-(4d+2)KP_1) 
+4dL_\mu L_\nu -2d(P_{1\mu}L_\nu+P_{1\nu}L_\mu)\\-2d(P_{2\mu}L_\nu+P_{2\nu}L_\mu)-g_{\mu \nu}(2dL^2-(4d+2)LP_2) \\
+(d-4)P_\mu P_\nu-g_{\mu \nu}(2d-3)P^2
\end{gathered}
\end{split}
\end{equation}
Remember that our integration variable was $S=K-P_1x$. In this substitution, keep in mind that $S_\mu =(s_\alpha,0)$ where $\alpha$ once again runs from 1 to d. Replacing $K_\mu =S_\mu +P_{1\mu}x$ and rewriting $f_{\mu \nu}$
\begin{equation}
\begin{split}
\begin{gathered}
f_{\mu \nu}=4d S_\mu S_\nu+4dx(S_\mu P_{1\nu}+S_\nu P_{1\mu}) +4d(S_\mu L_\nu+S_\nu L_\mu) \\-2d(P_{1\mu}S_\nu+P_{1\nu}S_\mu)-2d(P_{2\mu}S_\nu+P_{2\nu}S_\mu)-g_{\mu \nu}(2dS^2-(4d+2)SP_1) \\
+4dL_\mu L_\nu +(4dx-2d)(P_{1\mu}L_\nu+P_{1\nu}L_\mu)-2d(P_{2\mu}L_\nu+P_{2\nu}L_\mu)\\-g_{\mu \nu}(2dL^2-(4d+2)LP_2) \\
+(d-4)P_\mu P_\nu+4dx^2 P_{1\mu}P_{1\nu}-4dxP_{1\mu}P_{1\nu}-2dx(P_{2\mu}P_{1\nu}+P_{1\mu}P_{2\nu})\\-g_{\mu \nu}((2d-3)P^2+2dx^2P_1 ^2-(4d+2)xP_1^2)
\end{gathered}
\end{split}
\end{equation}

We will take integrals of the form 
\begin{equation}
\int d^d s \; e^{-\alpha s^2}(4d S_\mu S_\nu-g_{\mu \nu}(2dS^2)\text{+odd terms})
\label{Sdependent}
\end{equation}
\begin{equation}
\int d^d s \; e^{-\alpha s^2} \text{(the rest of $f_{\mu \nu}$)}
\end{equation}
where odd terms will not contribute.  Let us investigate \eqref{Sdependent} closer. For the first d terms we have (recall that D=d+2):
\begin{equation}
4d S_\mu S_\nu-g_{\mu \nu}(2dS^2)\rightarrow 2ds_1 ^2-2ds_2^2-2ds_3^2-....-2ds_d^2 
\label{Sexpanded}
\end{equation}
can be written for the first element. For each term in \eqref{Sexpanded} for instance, we will evaluate
\begin{equation}
2d\int ds_1 s_!^2 e^{-\alpha s_1^2}\int ds^{d-1}e^{-\alpha s^2}=\frac{2d\sqrt{\pi}}{2\alpha ^{3/2}}\left(\sqrt{\frac{\pi}{\alpha}}\right)^{d-1}=\frac{d}{\alpha}(\frac{\pi}{\alpha})^{d/2}
\label{Sintegral}
\end{equation}
We wrote the expression in \eqref{Sintegral} like this because we will take the parenthesis of $(\frac{\pi}{\alpha})^{d/2}$ together with the rest of $f_{\mu \nu}$ 
\begin{equation}
\int d^ds e^{-\alpha s^2}=\left(\frac{\pi}{\alpha}\right)^{d/2}
\end{equation}
Recall that in \eqref{Sintegral} the prefactor $\frac{d}{\alpha}$ is obtained for each term in \eqref{Sexpanded} therefore each element of $g_{\mu \nu}$ will result in the following constant
\begin{equation}
   -\frac{d}{\alpha}(d-2)g_{\mu \nu} =-\frac{d^2}{\alpha}g_{\mu \nu}+\frac{2d}{\alpha}g_{\mu \nu}
   \label{firstd-2factor}
\end{equation}
For the last two elements in \eqref{Sdependent}, we have the following factor:
\begin{eqnarray}
-\frac{d^2}{\alpha}g_{\mu \nu}
\label{lasttwofactor}
\end{eqnarray}
Expressing these in a cleaner a fashion and denoting with $r_{\mu \nu}$:
\[ \begin{cases} 
      -\frac{d^2}{\alpha}g_{\mu \nu}+\frac{2d}{\alpha}g_{\mu \nu} & (\mu , \nu)\leq (d,d) \\
     -\frac{d^2}{\alpha}g_{\mu \nu} &(d,d) < (\mu , \nu) \leq (D,D)
   \end{cases}
\]

Writing the cosine term in the polarization tensor in terms of exponentials and gathering all we had, we can write the polarization tensor
\begin{equation}
\frac{g^2 \pi^{d/2}}{2(2\pi)^2(2\pi)^d}\sum_{\epsilon \pm 1}\sum_{\vec{l}\epsilon Z^2}\int _0 ^1 dx \int _0 ^\infty \frac{d\alpha}{\alpha^{d/2-1}}e^{-\alpha (L^2-2LP_2x+P_1^2x(1-x)+P_2^2x)}e^{2\pi i \theta \epsilon l_i n_j \epsilon_{ij}}f_{\mu \nu}
\end{equation}
with $f_{\mu \nu}$ is given by
\begin{equation}
\begin{split}
\begin{gathered}
f_{\mu \nu}=4dL_\mu L_\nu-g_{\mu \nu}(2dL^2)+(4dx-2d)(P_{1\mu}L_{\nu}+P_{1\nu}L_{\mu})-2d(P_{2\mu}L_\nu+P_{2\nu}L_\mu)\\+g_{\mu \nu}(4d+2)LP_2 
(d-4)P_\mu P_\nu +4dx^2 P_{1\mu}P_{1\nu}-4dx(P_{1\mu}P_{1\nu})\\-2dx(P_{2\mu}P_{1\nu}+P_{1\mu}P_{2\nu}) -g_{\mu \nu}((2d-3)P^2+2dx^2P_1^2-(4d+2)xP_1^2)\text{+ $r_{\mu \nu}$}
\end{gathered}
\end{split}
\end{equation}

Now, we will apply Poisson resummation for this expression. For the Poisson resummation, we will apply the following:
\begin{eqnarray}
\sum_{l} f(l)=\sum_{K_1,K_2}\int dl_1 dl_2 f(l_1,l_2)e^{i2\pi K_1 l_1}e^{i2\pi K_2 l_2}
\end{eqnarray}
When we apply Poisson resummation to the $f_{\mu \nu}$, we get a constant factor similar to equation \eqref{firstd-2factor} and \eqref{lasttwofactor}. We will have the following
 \[ \begin{cases} 
      -\frac{2d}{\alpha}g_{\mu \nu} & (\mu , \nu)\leq (d,d) \\
     0 &(d,d) < (\mu , \nu) \leq (D,D)
   \end{cases}
\]   
coming from Poisson resummations of the terms such as $l_1 ^2$ and $l_2 ^2$. In overall, combined with \eqref{firstd-2factor} and \eqref{lasttwofactor} the overall constant factor becomes 
\begin{equation}
    -\frac{d^2}{\alpha}g_{\mu \nu}
\end{equation}
Therefore, after the Poisson resummation, we obtain the following expression:
\begin{equation}
    \frac{g^2 }{2(4\pi)^{d/2+1}}\sum_{\epsilon \pm 1}\sum_{\vec{K}\epsilon Z^2}\int _0 ^1 dx \int _0 ^\infty \frac{d\alpha}{\alpha^{d/2}}e^{-(\frac{\pi ^2 G^2}{\alpha}+\alpha P^2x(1-x))}e^{-2\pi i x \epsilon G.P } f_{\mu \nu}
\end{equation}
where we defined $G_\mu =(0,K_i -\epsilon _{ij}\theta n_j)$ and $K_i$'s are momenta after Fourier transformation. Moreover, $f_{\mu \nu}$ is written as 
\begin{equation}
\begin{split}
[-\frac{\pi ^2}{\alpha ^2}(4dG_\mu G_\nu-2dg_{\mu \nu}(G^2-\frac{d \alpha}{2\pi^2}))]\\+
   [\frac{-i\pi \epsilon}{\alpha}[(4dx-2d)((P_\mu G_v +P_\nu G_\mu)+g_{\mu \nu}(-4dx+4d+2)G.P]]\\+(4dx(x-1)+d-4)P_\mu P_\nu-g_{\mu \nu}((2d-3)+2dx^2-(4d+2)x)P^2
\end{split}
\end{equation}

One can utilize properties of the Bessel function to show that
\begin{equation}
K_{v+1}(z)-K_{v-1}(z)=\frac{2v}{z}K_v(z)
\end{equation}
From this relation, one can show that
\begin{eqnarray}
\begin{gathered}
\int _0 ^\infty \frac{d\alpha}{\alpha ^{d/2+2}}e^{-z^2\alpha+\frac{1}{\alpha}}(1-\frac{d\alpha}{2})=    z^{d/2+1}\int _0 ^\infty \frac{d\alpha}{\alpha ^{d/2+2}}e^{-z(\alpha+\frac{1}{\alpha})}(1-\frac{d\alpha}{2z}) \\
=2z^{d/2+1}[K_{d/2+1}(2z)-\frac{d}{2z}K_{d/2}(2z)]=2z^{d/2+1}K_{d/2-1}(2z)\\=z^2\int _0 ^\infty \frac{d\alpha}{\alpha ^{d/2}}e^{-z(\alpha+\frac{1}{\alpha})}
\label{Besseltrick}
\end{gathered}
\end{eqnarray}
Now we will use this trick to show that the $g_{\mu \nu}$ term which has $G^2$ can be written as a $P^2$ term. Let us show this:
\begin{equation}
2dg_{\mu \nu} \pi ^2 G^2 \int _0 ^\infty \frac{d\alpha}{\alpha ^{d/2+2}}e^{-\alpha P^2x(1-x)-\frac{\pi ^2 G^2}{\alpha}}(1-\frac{d\alpha}{2\pi ^2 G^2})
\end{equation}
Substituting $\alpha \rightarrow \alpha \pi ^2 G^2$
\begin{equation}
=\frac{2d g_{\mu \nu} \pi ^2 G^2}{(\pi ^2 G^2)^{d/2+1}}\int _0 ^\infty \frac{d\alpha}{\alpha ^{d/2+2}}e^{-z^2\alpha-\frac{1}{\alpha}}(1-\frac{d\alpha}{2})
\end{equation}
where we used the short notation $z^2=\pi ^2 G^2 P^2x(1-x)$. Remembering \eqref{Besseltrick} one can show that
\begin{equation}
=\frac{2d g_{\mu \nu} \pi ^2 G^2}{(\pi ^2 G^2)^{d/2+1}} z^2 \int _0 ^\infty \frac{d\alpha}{\alpha ^{d/2}}e^{-z^2\alpha-\frac{1}{\alpha}}
\end{equation}
Reverting back with $\alpha \rightarrow \frac{\alpha}{\pi ^2 G^2}$,
\begin{equation}
=2dg_{\mu \nu}P^2x(1-x) \int _0 ^\infty \frac{d\alpha}{\alpha ^{d/2}}e^{-\alpha P^2x(1-x)-\frac{\pi ^2 G^2}{\alpha}}
\end{equation}
Therefore instead of $\frac{2dg_{\mu \nu}}{\alpha ^2}(\pi ^2 G^2 -\frac{d\alpha}{2})$ we can use $2dg_{\mu \nu}P^2 x(1-x)$. With this, we were able to get rid of a term that can create problems with Ward identities.

Currently, we have
\begin{eqnarray}
\frac{g^2 }{(4\pi)^{d/2+1}}\sum_{\vec{K}\epsilon Z^2}\int _0 ^1 dx \int _0 ^\infty \frac{d\alpha}{\alpha^{d/2}}e^{-(\frac{\pi ^2 G^2}{\alpha}+\alpha P^2x(1-x))}\cos{(2\pi x G.P )} f_{\mu \nu}
\end{eqnarray}
with $f_{\mu \nu}$ given by
\begin{equation}
\begin{split}
[-\frac{\pi ^2}{\alpha ^2}4dG_\mu G_\nu]\\+
   [\frac{-i\pi \epsilon}{\alpha}[(4dx-2d)((P_\mu G_v +P_\nu G_\mu)+g_{\mu \nu}(-4dx+4d+2)G.P]]\\+(4dx(x-1)+d-4)P_\mu P_\nu-g_{\mu \nu}((4dx(x-1)+d-4)+d+1-2dx-2x)P^2
\end{split}
\end{equation}

We denoted the loop momenta after Poisson resummation with the letter K to be able to cleanly take the Fourier transform. In the end, the letter K represents the loop momenta on torus therefore now, we plan to switch back to the original variable which was denoted with the letter l.Therefore, we will write $G_{\mu}=(0,l_i-\theta \epsilon_{ij}n_j)$. \par 
Instead of summing over all l values, we will keep the minimum of it denoted by $\vec{l_0}$ such that $G_\mu$ becomes minimum. We denote the minimum as $b_i=(l_0)_i-\theta \epsilon_{ij}n_j$. The first order correction to the self-energy will be mainly dominated by the $G^2$ term and $P.G$ terms are taken as small \cite{Guralnik}.  Let us simply calculate the one-loop potential using the following identities:
\begin{align}
\int _0 ^\infty \frac{d\alpha}{\alpha ^{d/2+2}} e^{-\alpha P^2 x(1-x)-\frac{\pi ^2 G^2}{\alpha ^2}}=2\left(\frac{\sqrt{P^2x(1-x)}}{\pi |G|}   \right)^{d/2+1}K_{d/2+1}(2\pi |G|\sqrt{P^2x(1-x)})
\label{Integralexpansion}
\end{align}
and recalling the expansion of the Bessel function
\begin{equation}
K_{d/2+1}(2z)=\frac{\Gamma(d/2+1)}{2z^{d/2+1}}-\frac{\Gamma(d/2)}{2z^{d/2-1}}+...
\label{Besselexpansion}
\end{equation}
rewriting the \eqref{Integralexpansion} up to the first order in the expansion
\begin{eqnarray}
2\left(\frac{\sqrt{P^2x(1-x)}}{\pi |G|}   \right)^{d/2+1}K_{d/2+1}(2\pi |G|\sqrt{P^2x(1-x)}=\frac{\Gamma(d/2+1)}{\pi ^{d+2} |G|^{d+2}}
\end{eqnarray}
Integration of cos term in the limit $\vec{n}.\vec{b}\rightarrow 0$ gives a delta function. We find the final result
\begin{equation}
\Pi _{\mu \nu} ^1=-\frac{\Gamma(d/2+1)d g^2 b_\mu b_
\nu}{2^d \pi^{3d/2+1}{|\vec{b}|^{d+2}}}
\label{tachyonic}
\end{equation}
The quantum inverse propagator, from which the dispersion relation is generated, is given by
\begin{eqnarray}
\Gamma_{\mu \nu} ^{(1)}=(w^2-n^2)g_{\mu \nu}-\Pi_{\mu \nu}
\label{quantuminversepropagator}
\end{eqnarray}
Therefore,
\begin{eqnarray}
w^2=\frac{n^2}{R^2}-\frac{dg^2\Gamma (d/2+1)}{(2R)^d \pi ^{3/2d+1}} \frac{1}{|\vec{b}|^d}
\label{instability}
\end{eqnarray}
where R dependence is reintroduced. One can see that for D=2,d=0 potential's value is 0 as expected since gluons do not have any propagating degrees of freedom. At sufficiently large N and small $\vec{b}$, \eqref{instability} leads to a tachyonic instability. This is a rare example where quantum corrections can surpass the tree-level term contribution.

\section{Fixing the Instability}
\label{sec: Fixing the Instability}
In this section, similar to the gauge field part, we compute the adjoint fermion-gluon diagram and show that it eliminates the instability. We will include the following lagrangian on top of Yang-Mills \cite{Armoni1}
\begin{equation}
L_{fermion}=\overline{\psi}i\slashed{\partial}\ast \psi+A_\mu \ast \overline{\psi}\gamma^\mu \ast \psi -\overline{\psi}\ast A_\mu \gamma^\mu \ast \psi
\end{equation}
with the contribution of the diagram for the massless adjoint fermions will be given for the U(1) gauge group and since the gauge group is U(1), it puts constraints on the matter field representations, only fundamental,anti-fundamental, and adjoint representations are possible \cite{Chaichian}.
\begin{figure} [h]
\begin{center}
\includegraphics[width=0.4\textwidth]{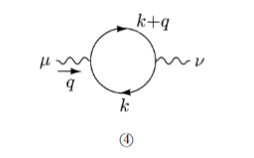} 
\end{center}
\caption{Fermion-photon loop diagram from \cite{PeskinSchroeder199508}. Wavy lines belong to photon whereas solid lines are fermions.}
\label{fig:stage}
\end{figure}

 The usual diagram from \cite{PeskinSchroeder199508} reads as
\begin{equation}
\textcircled{4}=- g^2\int \frac{d^D k}{(2\pi)^D} Tr[\gamma^\mu \frac{1}{\slashed{k}} \gamma^\nu \frac{1}{\slashed{k}+\slashed{q}}]
\end{equation}
One can write this 
\begin{equation}
\textcircled{4}=- g^2\int \frac{d^D k}{(2\pi)^D} Tr[\gamma^\mu \frac{\slashed{k}}{k^2} \gamma^\nu \frac{\slashed{k}+\slashed{q}}{(k+q)^2}]
\label{fermiondiagram}
\end{equation}
Using the trace identities of gamma matrices
\begin{equation}
Tr[\gamma ^\mu \slashed{k}\gamma ^\nu (\slashed{k}+\slashed{q})]=k_s (k+q)_\sigma Tr[\gamma ^\mu \gamma ^s \gamma ^\nu \gamma ^\sigma ]=4 k_s (k+q)_\sigma (g^{\mu s}g^{\nu \sigma}-g^{\mu \nu}g^{s \sigma}+g^{\mu \sigma}g^{\nu s}) 
\end{equation}
Then we can rewrite the \eqref{fermiondiagram} as
\begin{equation}
\textcircled{4}=- 4g^2\int \frac{d^D k}{(2\pi)^D} \frac{1}{k^2 (k+q)^2} [k^\mu (k+q)^\nu +k^\nu (k+q)^\mu -g^{\mu \nu}k (k+q)]
\end{equation}
Now, we do the following: writing momenta's in Guralnik's notation, introducing the nonplanarity \cite{Minwalla,Armoni1} and summing over loop momenta on $T^2$ torus 
\begin{equation}
\frac{-4g^2}{(2\pi R)^2 (2\pi)^d} \sum_{\vec{l}\epsilon Z^2}\int \; d^dk \frac{[K^\mu (K-P)^\nu + K^\nu (K-P)^\mu-g^{\mu \nu}K(K-P)]}{K^2 (K-P)^2} e^{2\pi i \theta l_j n_k \epsilon_{jk}}
\label{fermionnoncommutative}
\end{equation}
where define $K_\mu =(k_\alpha ,l_i)$ and $P_\mu=(p_\alpha ,n_i)$. Similar to gauge fields we will suppress R by taking $R=1$ until the end of the calculation 

\par 
Notice that when $\theta \rightarrow 0$ limit is taken in \eqref{fermionnoncommutative}, it recovers the normal diagram \cite{Armoni1}. The extra minus will again be incorporated in the quantum corrected inverse propagator. \par 
Applying the same procedure given in the gauge field part, introducing Feynman parameters, Schwinger parameters, separating momenta, integrating out momenta in noncompact directions, and Poisson resummation, we obtain the following polarization tensor:

\begin{equation}
\frac{-4g^2}{2^D \pi ^{d/2+1}} \sum_{\vec{l}\epsilon Z^2}\int _0 ^1 dx \int _0 ^\infty \frac{d\alpha}{\alpha ^{d/2}} \;  e^{-\alpha (P^2 x(1-x)-\frac{\pi ^2 G^2}{\alpha}}e^{-2\pi i x P.G}  f'_{\mu \nu}
\end{equation}
where we used $P_2.G=P.G$, returned to the original loop momentum $G_{\mu}=(0,l_i-\theta \epsilon_{ij}n_j)$ and we defined
\begin{equation}
\begin{split}
f'_{\mu \nu}=-\frac{\pi ^2}{\alpha ^2}(2G_\mu G_\nu-g_{\mu \nu}(G^2-\frac{d \alpha}{2\pi ^2})) \\
-\frac{i \pi}{\alpha}(2x-1)((G_\mu P_\nu+G_\nu P_\mu)-g_{\mu \nu}P.G) \\
+x(x-1)(2P_\mu P_\nu-g_{\mu \nu}P^2)
\end{split}
\end{equation}
Once again using the properties of the Bessel function, we can write $\frac{g_{\mu \nu}}{\alpha ^2}(\pi ^2 G^2 -\frac{d\alpha}{2})=2dg_{\mu \nu}P^2 x(1-x)$. 
Expanding $G_{\mu}=(0,l_i-\theta \epsilon_{ij}n_j)$ around its minimum value, denoting by $\vec{b}=(0,l_{(0)i}-\theta \epsilon_{ij}n_j)$ and using the same equations given in \eqref{Integralexpansion} and \eqref{Besselexpansion}, taking $P.G$ and $P^2$ terms small \cite{Guralnik}, we find for $n_f$ adjoint fermions
\begin{equation}
\Pi^1=+n_f \frac{8 \Gamma(d/2+1)g^2}{2^D \pi ^{3d/2+1}|\vec{b}|^d}
\end{equation}
Remember that for the gauge field, we found
\begin{equation}
\Pi^1=- \frac{d \Gamma(d/2+1)g^2}{2^d \pi ^{3d/2+1}|\vec{b}|^d}
\end{equation}
It is straightforward to show that for $D=4$, $d=2$ and we replaced back the R dependence
\begin{equation}
\Pi^1=(n_f-1)\frac{g^2}{2\pi ^4 R^2 |\vec{b}|^2}
\end{equation}
and we reached to the stabilization.

\section{Conclusions and Future Directions}
\label{sec:conclusionsandfuturedirections} 
In this work, we investigated the $\Z_N \times \Z_N$ symmetry breaking on $T^2 \times \R^2$ compactification of the Yang-Mills theory with twisted boundary conditions. Center symmetry realization exhibits an interesting 
non-commutativity of limits between small $L$ and large-N: For fixed N, it is stable for sufficiently small $L$, however, at fixed $L$, 
vacuum is destabilized by the quantum corrections at sufficiently large N.   Using Morita equivalence on torus, we converted the problem to the tachyonic instability seen in noncommutative theories. Later, working on the noncommutative side, we discuss ways to overcome the instability. Similar to theories on $\R^3 \times S^1$, the inclusion of adjoint fermions work changes quantum fluctuations and it favors stabilization of the 
center symmetry. \par 
There are some problems to pursue further. One is to understand 
 what would center vortices correspond to on the noncommutative side. 
 For a complete understanding of confinement on the noncommutative side, this is a necessary step to know. Moreover, it is curious to study Seiberg-Witten on the same setup. Especially, on $T^2 \times T^2$ compactification with 't Hooft flux turned on at each $T^2$ fractional topological charge object can be observed. It is also interesting to use Poisson duality, to show how center vortex theory continuously connects to the Seiberg Witten theory as already conducted for theories on $\R^3 \times S^1$ \cite{Poppitz:2011wy} and noncommutative analogies for these.  In particular, it would be interesting to understand the 
 structure of center-vortices in ${\cal N}=2$  Seiberg-Witten theory and its small mass deformation to ${\cal N}=1$. Since the former is non-confining on $\mathbb R^4$, and the latter is confining, we expect an interesting change in the structure of the center-vortices on 
 $\mathbb R^2 \times T^2$ with $n_{12}$ units of 't Hooft flux.  This investigation can be performed by combining ideas along the lines of \cite{Hayashi:2024psa, Hayashi:2024yjc, Guvendik:2024umd, Poppitz:2011wy}.

\acknowledgments
C.G. thanks his academic advisor  Mithat Unsal and Thomas Schaefer  
for many helpful discussions. The work is supported by the U.S. Department of Energy, Office of Science, Office of Nuclear Physics under Award Number DE-FG02-03ER41260.

\bibliographystyle{utphys}
\bibliography{QFT-Mithat,refs}
\end{document}